\documentclass{PoS}

\title{Jet precession in the active nucleus of M81. Ongoing VLBI monitoring.}

\ShortTitle{Monitoring the jet precession in the AGN of M81}

\author{\speaker{I. Marti-Vidal}\\
        Onsala Space Observatory (Chalmers Univ. of Technology, Sweden)\\
        E-mail: \email{ivan.marti-vidal@chalmers.se}}

\author{J.M. Marcaide\\
        Dpt. Astronomia i Astrof\'isica (Univ. Valencia, Spain)\\
        E-mail: \email{j.m.marcaide@uv.es}}
        
\author{A. Alberdi\\
        Instituto de Astrof\'isica de Andaluc\'ia, CSIC (Granada, Spain)\\
        E-mail: \email{antxon@iaa.es}}        
        
\author{A. Brunthaler\\
        Max-Planck-Institut f\"ur Radioastronomie (Bonn, Germany)\\
        E-mail: \email{brunthal@mpifr-bonn.mpg.de}}

\abstract{In a recent publication, we reported results of a multi-frequency VLBI campaign of observations of the Active Galactic Nucleus (AGN) in galaxy M\,81, phase-referenced to the supernova SN\,1993J. We were able to extract precise information on the relative astrometry of the AGN radio emission at different epochs and frequencies. We found strong evidence of precession in the AGN jet (i.e., a systematic evolution in the jet inclination at each frequency) coupled to changes in the overall flux density at the different frequencies. In these proceedings, we summarise the main contents of our previous publication and we report on (preliminary) new results from our follow-up VLBI observations, now phase-referenced to the young supernova SN2008iz. We also briefly discuss how these results match the picture of our previously-reported precession model.}

\FullConference{11th European VLBI Network Symposium \& Users Meeting,\\
		October 9-12, 2012\\
		Bordeaux, France}

\begin{document}

\section{Introduction}

The galaxy M\,81 (NGC\,3031), located at a distance of 3.63$\pm$0.34\,Mpc (Freedman et al. \cite{Freedman1994}), hosts the next closest active galactic nucleus (AGN) after Centaurus A. Its high declination (J2000.0 coordinates $\alpha = 09^\textrm{h} 55^\textrm{m} 33.173^\textrm{s}$ and $\delta =
69^\circ 03' 55.062''$) also makes the AGN in M\,81 (hereafter M\,81*) a perfect target for global VLBI observations using the most sensitive radio telescopes in the Northern Hemisphere.

The explosion of the radio-luminous supernova SN\,1993J in M\,81, which happened around 28 March 1993 (e.g., Weiler et al. \cite{Weiler2007}), triggered an intense campaign of VLBI observations in which M\,81* was selected as the phase calibrator of the SN\,1993J visibilities (e.g., Marcaide et al. \cite{Marcaide1995},\cite{Marcaide1997}; Bartel et al. \cite{Bartel2002}; Bietenholz et al. \cite{BietenIII}; Marcaide et al. \cite{Marcaide2009}; Marti-Vidal et al. \cite{PaperI},\cite{PaperII}). These observations allowed us to study, as a byproduct, the
structure and evolution of M\,81* at different frequencies (Marti-Vidal et al. \cite{Marti2011}). Since the central position of the radio-emitting shell of a supernova does not depend on the observing frequency, this is the ideal study case for multi-frequency astrometry of an AGN jet.

Results for some of the M\,81* observations at 8.4\,GHz between years 1993 and 1997 are reported in Bietenholz et al. \cite{Bieten2000}, and results for the astrometry analysis of a subset of observations at all frequencies between years 1993 and 2002 were reported in Bietenholz et al. \cite{Bieten2004}. These authors concluded that the proper motion of M\,81*, relative to the shell center of SN 1993J, was consistent with zero. The contributions of galactic rotation and/or any peculiar motion of the SN\,1993J progenitor were also discarded to contribute to a detectable proper motion of M\,81* relative to the supernova. These authors also reported a frequency-dependent shift of the peak of brightness of M\,81*.

In Marti-Vidal et al. \cite{Marti2011}, we reported the analysis of the combined dataset of all the VLBI observations of SN\,1993J at frequencies from 1.4 to 8.4\,GHz, phase-referenced to M\,81*, that were publicly available at the time of the analysis. The time sampling of this combined dataset\footnote{Observations from a project lead by N. Bartel at Univ. York in Toronto (Canada) and observations from a project lead by J.M. Marcaide at Univ. Valencia (Spain)} is more dense and extended than those of the results previously reported in Bietenholz et al. \cite{Bieten2000} and Bietenholz et al. \cite{Bieten2004} (see Fig. \ref{Fig2}). Hence, any systematic evolution in the source position should be more clearly revealed in our analysis, as it is, indeed, the case (see Sect. \ref{Sec11}).

\section{Summary of Previous Results}
\label{Sec11}

In Marti-Vidal et al. \cite{Marti2011}, we confirmed the main results reported in Bietenholz et al. \cite{Bieten2000} and Bietenholz et al. \cite{Bieten2004}; on the one hand, the position of the intensity peak in the AGN jet (namely, its {\em core}) depends on the observing frequency, and is shifted along the jet towards larger distances from the jet base (i.e., the AGN central {\em engine}) as we decrease the observing frequency. On the other hand, the core region is slightly resolved at all frequencies, and its size increases with decreasing frequency. 

The chromatic shift in the position of the peak intensity of an AGN jet (also known as the {\em core-shift effect}) was serendipitously 
discovered in the pair of quasars 1038+528\,A-B (Marcaide \cite{MarcaideTesis}) and has been extensively studied in many sources during the last years (e.g., Kovalev et al. \cite{Kovalev}). This apparent shift in the source position is caused by Synchrotron Self-absorption (SSA) of the jet plasma, due to the strong magnetic fields and the large particle densities (and energy gradients) close to the jet base (Blandford \& K\"onigl \cite{Blandford1979}).

An increasing source size with decreasing frequency, coupled to the core-shift effect described in the previous paragraph, maps into a direct measurement of the opening angle in the jet of M\,81*, which of course has to be corrected for geometric projection effects (e.g., Sect. 5.4 in Marti-Vidal et al. \cite{Marti2011}). 

Besides the confirmation of the main results reported by other authors, the larger amount of data available allowed us to to better constrain proper motions and to estimate core-shifts with a higher precision. We were able to constrain the proper motion of the AGN {\em core} (relative to the frame of SN\,1993J, located in the same galaxy), with a precision of a few $\mu$as per year (depending on frequency). We were also able to determine the {\em normalised core-shift}, $\Omega$ (see Lobanov \cite{Lobanov1998}), from a combined fit of all the available shifts for the different frequency pairs, with an uncertainty of only 11\%. With this precise estimate of the core shift, we could derive the coordinates of the jet base of M\,81* with precisions ranging from 20$\mu$as (at 8.4\,GHz) to 100$\mu$as (at 1.7\,GHz). Our results were indeed compatible with the estimates given in Bietenholz et al. \cite{Bieten2004}, who used a completely different approach to derive the coordinates of the jet base. Last but not least, we could also estimate the mass of the central black hole, by assuming a strongly-magnetised BH scenario (Kardashev \cite{Kardashev1995}). The resulting black-hole mass ($2\times10^7$\,M$_\odot$) is in agreement with other independent mass estimates (coming from spectroscopic observations of the central rotating star disk and from the stellar-velocity dispersion in the bulge) and 
makes M\,81* an interesting intermediate object between Sgr\,A* and the more powerful AGN.

\subsection{The precessing jet in M\,81*}
\label{Sec111}

\begin{figure}[ht!]
\centering
\includegraphics[width=15cm]{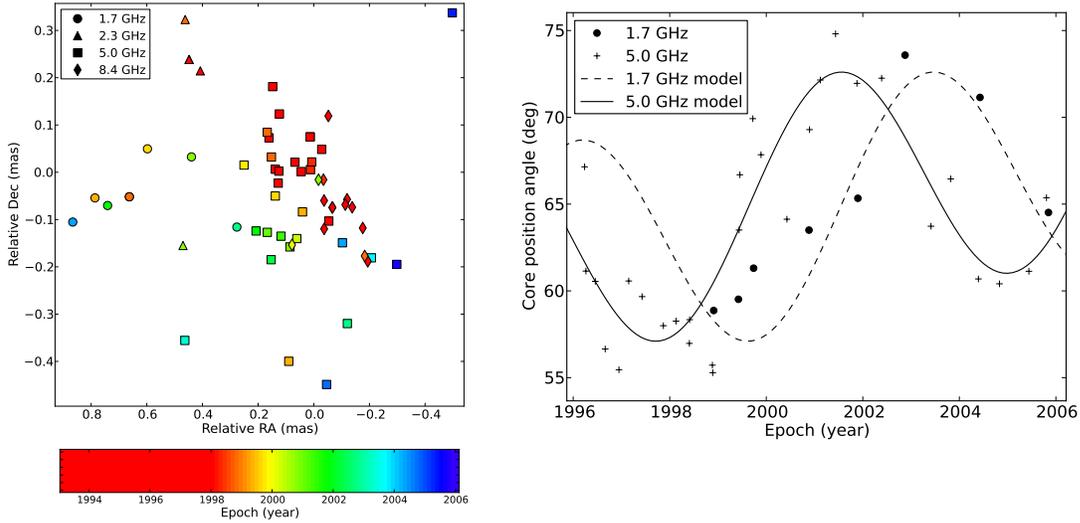}
\caption{Left, position of the peak intensity of M\,81* for each frequency (different symbols) and epoch (different colors). The {\em core-shift} effect can be appreciated from the gathering of the different symbols around different coordinates, whereas the jet precession can be appreciated from the gathering of symbols with similar colors. Right, position angle of the core at 5\,GHz and 1.7\,GHz as a function of time. A simple sinusoidal precession model is over-plotted.}
\label{Fig1}
\end{figure}

There are strong signatures of a systematic change in the orientation of the jet, both from the phase-referencing to SN\,1993J (Fig. \ref{Fig1}, left) and from the evolving inclination of the core, as estimated from model-fitting to the visibilities\footnote{The position angle of the core major axis is almost independent of the underlying model used in the fit (disc, Gaussian, sphere, etc.), as long as the source is only marginally resolved. In such a case, we are only able to sample the first lobe of the visibility function, which has the shape of a paraboloid with minor axis {\em aligned} to the major axis of the source intensity distribution. Given that {\em any} symmetric model of a small source has a main lobe with a paraboloidal shape, the position angle of the minor axis in visibility space (i.e., the position angle of the major axis of the source) shall be independent of the specific model used to estimate it.} (Fig. \ref{Fig1}, right).

This change in the orientation of the jet seems to be coupled to long-term changes in the VLBI flux density at the different frequencies. This is probably due to jet bending effects or to inhomogeneities in the jet path as it changes direction. In Fig. \ref{Fig3}, we show the large flare (4 years long) reported in Marti-Vidal \cite{Marti2011}, which happens roughly when the jet increases its position angle from North to East.

\section{New Observations}

The serendipitous discovery of supernova SN2008iz in M\,82 (Brunthaler et al. \cite{Brunt1}), targeted an intense VLBI monitoring at different frequencies, to characterise its structure evolution and expansion (e.g., Brunthaler et al. \cite{Brunt2},\cite{Brunt3}).

\begin{figure}[ht!]
\centering
\includegraphics[width=15cm]{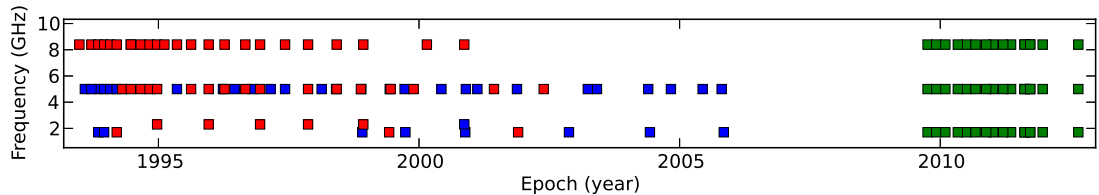}
\caption{Time sampling of our VLBI monitoring. Red points are epochs also reported in Bietenholz et al. \cite{Bieten2004}; blue points are epochs from the project led by J.M. Marcaide; green points are the new observations, phase-referenced to SN\,2008iz.}
\label{Fig2}
\end{figure}

Given that M\,81* was again selected as phase calibrator,
we can use these new observations to continue our study of the evolving jet in M\,81*. We show in Fig. \ref{Fig2} the time distribution of all the epochs used so far in our analysis. These are those previously reported in Bietenholz et al. \cite{Bieten2004} (red), Marti-Vidal et al. \cite{Marti2011} (red and blue), and the new observations (green), phase-referenced to SN\,2008iz and first reported in these proceedings, that will be discussed in deeper detail elsewhere (Marti-Vidal et al. in preparation).

\section{New Preliminary Results}

We have generated hybrid images of M\,81* for all the new epochs of observation of SN\,2008iz (i.e., those corresponding to the green points shown in Fig. \ref{Fig2}), and have performed model-fitting to the visibilities in a subset of epochs. The model structure and the fitting algorithm used are described in Marti-Vidal et al. \cite{Marti2011} (see their Sect. 3.1).

\begin{figure}[ht!]
\centering
\includegraphics[width=10cm]{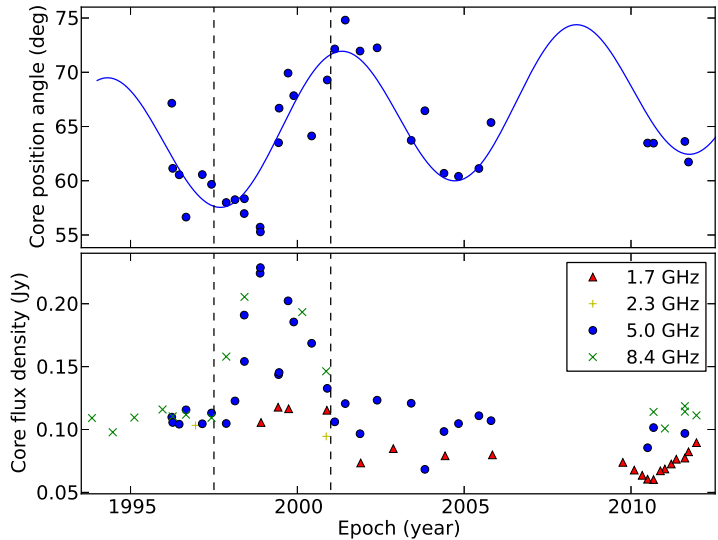}
\caption{Top, position angle of the M\,81* core vs. time. Bottom, lightcurve of the M\,81* core, obtained from the hybrid imaging of all our available VLBI epochs.}
\label{Fig3}
\end{figure}

We show in Fig. \ref{Fig3} the preliminary results (overall VLBI flux density of the core at bottom; core inclination at top) for these new epochs, together with the results reported in Marti-Vidal et al. \cite{Marti2011}. Since we are still working on the relative astrometry between M\,81* and SN\,2008iz, it is not possible to show information on the core-shift, for the new epochs, in these proceedings. In any case, it can be seen that the position angle at 5\,GHz for the new epochs matches well the prediction by the simple model reported in Marti-Vidal et al. \cite{Marti2011}. The difference between the position angles around year 2001 and year 2011 is obvious.

For clarity reasons, we have over-plotted two dashed lines marking the epochs of years 1997.5 and 2001. As reported in Marti-Vidal et al. \cite{Marti2011}, the long flare can be related to the increase of position angle of the core (i.e., when the core turns from North into East direction). In the case of a periodic precession, we would expect to see a second long flare roughly between years 2005 and 2008. Unfortunately, there is lack of observations just in this time range. We notice, though, that the core flux densities at 1.7\,GHz seem to reach a minimum short after year 2010, which corresponds to the time when the jet reaches its minimum position angle. The interpretation of these results in the frame of the precessing jet, together with the report of a more complete set of observations (now being reduced) will be discussed in a future publication.

\section{Summary}

We report new multi-frequency VLBI observations of M\,81*, phase-referenced to the supernova SN\,2008iz in M\,82, which will allow us to extend the analysis reported in Marti-Vidal et al. \cite{Marti2011}. We also report preliminary results on the flux-density evolution of the jet core at each frequency (8.4, 5.0, and 5.0\,GHz) and the position angle of the slightly-resolved elongated core. On the one hand, the new core position angles are in agreement with the extrapolation of the simple model reported in Marti-Vidal et al. \cite{Marti2011}. On the other hand, we lack observations in the time range when a second long flare would have been observed. Continued observations will allow us to finally confirm and improve, or reject, the periodicity of the precession model for the AGN jet in galaxy M\,81.

\subsection*{Acknowledgements}
This research has been partially supported by projects
AYA2009-13036-C02-01 and AYA2009-13036-C02-02 of the MICINN and by
grant PROMETEO 104/2009 of the Generalitat Valenciana.
A. B. was supported by
a Marie Curie Outgoing International Fellowship (FP7)
of the European Union (project number 275596).

\end{document}